EUROPEAN ORGANIZATION FOR NUCLEAR RESEARCH

# Experimental investigation of transverse spin asymmetries in $\mu$-p SIDIS processes: Collins asymmetries

The COMPASS Collaboration

**Abstract**

The COMPASS Collaboration at CERN has measured the transverse spin azimuthal asymmetry of charged hadrons produced in semi-inclusive deep inelastic scattering using a 160 GeV $\mu^+$ beam and a transversely polarised $NH_3$ target. The Collins asymmetry of the proton was extracted in the Bjorken $x$ range $0.003 < x < 0.7$. These new measurements confirm with higher accuracy previous measurements from the COMPASS and HERMES collaborations, which exhibit a definite effect in the valence quark region. The asymmetries for negative and positive hadrons are similar in magnitude and opposite in sign. They are compatible with model calculations in which the u-quark transversity is opposite in sign and somewhat larger than the d-quark transversity distribution function. The asymmetry is extracted as a function of Bjorken $x$, the relative hadron energy $z$ and the hadron transverse momentum $p_T^h$. The high statistics and quality of the data also allow for more detailed investigations of the dependence on the kinematic variables. These studies confirm the leading–twist nature of the Collins asymmetry.

**The COMPASS Collaboration**

C. Adolph[8], M.G. Alekseev[24], V.Yu. Alexakhin[7], Yu. Alexandrov[15,*], G.D. Alexeev[7], A. Amoroso[27], A.A. Antonov[7], A. Austregesilo[10,17], B. Badełek[30], F. Balestra[27], J. Barth[4], G. Baum[1], Y. Bedfer[22], J. Bernhard[13], R. Bertini[27], M. Bettinelli[16], K. Bicker[10,17], J. Bieling[4], R. Birsa[24], J. Bisplinghoff[3], P. Bordalo[12,a], F. Bradamante[25], C. Braun[8], A. Bravar[24], A. Bressan[25], E. Burtin[22], L. Capozza[22], M. Chiosso[27], S.U. Chung[17], A. Cicuttin[26], M.L. Crespo[26], S. Dalla Torre[24], S. Das[6], S.S. Dasgupta[6], S. Dasgupta[6], O.Yu. Denisov[28], L. Dhara[6], S.V. Donskov[21], N. Doshita[32], V. Duic[25], W. Dünnweber[16], M. Dziewiecki[31], A. Efremov[7], C. Elia[25], P.D. Eversheim[3], W. Eyrich[8], M. Faessler[16], A. Ferrero[22], A. Filin[21], M. Finger[19], M. Finger jr.[7], H. Fischer[9], C. Franco[12], N. du Fresne von Hohenesche[13,10], J.M. Friedrich[17], V. Frolov[10], R. Garfagnini[27], F. Gautheron[2], O.P. Gavrichtchouk[7], S. Gerassimov[15,17], R. Geyer[16], M. Giorgi[25], I. Gnesi[27], B. Gobbo[24], S. Goertz[4], S. Grabmüller[17], A. Grasso[27], B. Grube[17], R. Gushterski[7], A. Guskov[7], T. Guthörl[9], F. Haas[17], D. von Harrach[13], F.H. Heinsius[9], F. Herrmann[9], C. Heß[2], F. Hinterberger[3], N. Horikawa[18,b], Ch. Höppner[17], N. d'Hose[22], S. Ishimoto[32,c], O. Ivanov[7], Yu. Ivanshin[7], T. Iwata[32], R. Jahn[3], V. Jary[20], P. Jasinski[13], R. Joosten[3], E. Kabuß[13], D. Kang[13], B. Ketzer[17], G.V. Khaustov[21], Yu.A. Khokhlov[21], Yu. Kisselev[2], F. Klein[4], K. Klimaszewski[30], S. Koblitz[13], J.H. Koivuniemi[2], V.N. Kolosov[21], K. Kondo[32], K. Königsmann[9], I. Konorov[15,17], V.F. Konstantinov[21], A. Korzenev[22,d], A.M. Kotzinian[27], O. Kouznetsov[7,22], M. Krämer[17], Z.V. Kroumchtein[7], F. Kunne[22], K. Kurek[30], L. Lauser[9], A.A. Lednev[21], A. Lehmann[8], S. Levorato[25], J. Lichtenstadt[23], T. Liska[20], A. Maggiora[28], A. Magnon[22], N. Makke[22,25], G.K. Mallot[10], A. Mann[17], C. Marchand[22], A. Martin[25], J. Marzec[31], T. Matsuda[14], G. Meshcheryakov[7], W. Meyer[2], T. Michigami[32], Yu.V. Mikhailov[21], M.A. Moinester[23], A. Morreale[22], A. Mutter[9,13], A. Nagaytsev[7], T. Nagel[17], T. Negrini[9], F. Nerling[9], S. Neubert[17], D. Neyret[22], V.I. Nikolaenko[21], W.D. Nowak[9], A.S. Nunes[12], A.G. Olshevsky[7], M. Ostrick[13], A. Padee[31], R. Panknin[4], D. Panzieri[29], B. Parsamyan[27], S. Paul[17], E. Perevalova[7], G. Pesaro[25], D.V. Peshekhonov[7], G. Piragino[27], S. Platchkov[22], J. Pochodzalla[13], J. Polak[11,25], V.A. Polyakov[21], J. Pretz[4,e], M. Quaresma[12], C. Quintans[12], J.-F. Rajotte[16], S. Ramos[12,a], V. Rapatsky[7], G. Reicherz[2], A. Richter[8], E. Rocco[10], E. Rondio[30], N.S. Rossiyskaya[7], D.I. Ryabchikov[21], V.D. Samoylenko[21], A. Sandacz[30], M.G. Sapozhnikov[7], S. Sarkar[6], I.A. Savin[7], G. Sbrizzai[25], P. Schiavon[25], C. Schill[9], T. Schlüter[16], K. Schmidt[9], L. Schmitt[17,f], K. Schönning[10], S. Schopferer[9], M. Schott[10], W. Schröder[8], O.Yu. Shevchenko[7], L. Silva[12], L. Sinha[6], A.N. Sissakian[7,*], M. Slunecka[7], G.I. Smirnov[7], S. Sosio[27], F. Sozzi[24], A. Srnka[5], L. Steiger[24], M. Stolarski[12], M. Sulc[11], R. Sulej[30], H. Suzuki[32,b], P. Sznajder[30], S. Takekawa[28], J. Ter Wolbeek[9], S. Tessaro[24], F. Tessarotto[24], L.G. Tkatchev[7], S. Uhl[17], I. Uman[16], M. Vandenbroucke[22], M. Virius[20], N.V. Vlassov[7], L. Wang[2], M. Wilfert[13], R. Windmolders[4], W. Wiślicki[30], H. Wollny[9,22], K. Zaremba[31], M. Zavertyaev[15], E. Zemlyanichkina[7], M. Ziembicki[31], N. Zhuravlev[7] and A. Zvyagin[16]

[1] Universität Bielefeld, Fakultät für Physik, 33501 Bielefeld, Germany[g]
[2] Universität Bochum, Institut für Experimentalphysik, 44780 Bochum, Germany[g]
[3] Universität Bonn, Helmholtz-Institut für Strahlen- und Kernphysik, 53115 Bonn, Germany[g]
[4] Universität Bonn, Physikalisches Institut, 53115 Bonn, Germany[g]
[5] Institute of Scientific Instruments, AS CR, 61264 Brno, Czech Republic[h]
[6] Matrivani Institute of Experimental Research & Education, Calcutta-700 030, India[i]
[7] Joint Institute for Nuclear Research, 141980 Dubna, Moscow region, Russia[j]
[8] Universität Erlangen–Nürnberg, Physikalisches Institut, 91054 Erlangen, Germany[g]
[9] Universität Freiburg, Physikalisches Institut, 79104 Freiburg, Germany[g]
[10] CERN, 1211 Geneva 23, Switzerland
[11] Technical University in Liberec, 46117 Liberec, Czech Republic[h]
[12] LIP, 1000-149 Lisbon, Portugal[k]
[13] Universität Mainz, Institut für Kernphysik, 55099 Mainz, Germany[g]

[14] University of Miyazaki, Miyazaki 889-2192, Japan[l]
[15] Lebedev Physical Institute, 119991 Moscow, Russia
[16] Ludwig-Maximilians-Universität München, Department für Physik, 80799 Munich, Germany[f,l)]
[17] Technische Universität München, Physik Department, 85748 Garching, Germany[f,l)]
[18] Nagoya University, 464 Nagoya, Japan[l]
[19] Charles University in Prague, Faculty of Mathematics and Physics, 18000 Prague, Czech Republic[h]
[20] Czech Technical University in Prague, 16636 Prague, Czech Republic[h]
[21] State Research Center of the Russian Federation, Institute for High Energy Physics, 142281 Protvino, Russia
[22] CEA IRFU/SPhN Saclay, 91191 Gif-sur-Yvette, France
[23] Tel Aviv University, School of Physics and Astronomy, 69978 Tel Aviv, Israel[n]
[24] Trieste Section of INFN, 34127 Trieste, Italy
[25] University of Trieste, Department of Physics and Trieste Section of INFN, 34127 Trieste, Italy
[26] Abdus Salam ICTP and Trieste Section of INFN, 34127 Trieste, Italy
[27] University of Turin, Department of Physics and Torino Section of INFN, 10125 Turin, Italy
[28] Torino Section of INFN, 10125 Turin, Italy
[29] University of Eastern Piedmont, 15100 Alessandria, and Torino Section of INFN, 10125 Turin, Italy
[30] National Centre for Nuclear Research and University of Warsaw, 00-681 Warsaw, Poland[o]
[31] Warsaw University of Technology, Institute of Radioelectronics, 00-665 Warsaw, Poland[o]
[32] Yamagata University, Yamagata, 992-8510 Japan[l]
[a] Also at IST, Universidade Técnica de Lisboa, Lisbon, Portugal
[b] Also at Chubu University, Kasugai, Aichi, 487-8501 Japan[l]
[c] Also at KEK, 1-1 Oho, Tsukuba, Ibaraki, 305-0801 Japan
[d] On leave of absence from JINR Dubna
[e] present address: III. Physikalisches Institut, RWTH Aachen University, 52056 Aachen
[f] Also at GSI mbH, Planckstr. 1, D-64291 Darmstadt, Germany
[g] Supported by the German Bundesministerium für Bildung und Forschung
[h] Suppported by Czech Republic MEYS grants ME492 and LA242
[i] Supported by SAIL (CSR), Govt. of India
[j] Supported by CERN-RFBR grants 08-02-91009
[k] Supported by the Portuguese FCT - Fundação para a Ciência e Tecnologia, COMPETE and QREN, grants CERN/FP/109323/2009, CERN/FP/116376/2010 and CERN/FP/123600/2011
[l] Supported by the MEXT and the JSPS under the Grants No.18002006, No.20540299 and No.18540281; Daiko Foundation and Yamada Foundation
[m] Supported by the DFG cluster of excellence 'Origin and Structure of the Universe' (www.universe-cluster.de)
[n] Supported by the Israel Science Foundation, founded by the Israel Academy of Sciences and Humanities
[o] Supported by Ministry of Science and Higher Education grant 41/N-CERN/2007/0
[*] Deceased

Deep inelastic lepton-nucleon scattering (DIS) measurements as a tool to unveil the structure of the nucleon started in the late 60's at the Stanford Linear Accelerator (SLAC), when for the first time a high energy electron accelerator became available. A wealth of *eN* scattering data was collected, and eventually it became clear that scattering at large momentum transfer could be interpreted as elastic scattering off the nucleon constituents, the "partons". From the dependence of the cross–section on the energy and the momentum transfered to the nucleon it was possible to identify the charged partons with the earlier postulated quarks. In the subsequent years the SLAC energy was gradually increased. Higher energy experiments could be performed at CERN and FNAL using muon beams from $\pi$ and K decays, and at the HERA *ep* collider. The high beam energies allowed measurements at larger values of $Q^2$, the square of the four-momentum transfered to the nucleon, and at smaller values of $x$, the fraction of the nucleon momentum carried by the parton. All these data in combination with the neutrino-nucleon data eventually allowed the extraction of the nucleon parton distribution functions (PDFs) and their $Q^2$ dependence, in particular of the functions $q(x,Q^2)$ (or $f_1^q(x,Q^2)$), which describe the number density of quarks of flavour $q$.

In a second generation of DIS experiments, polarised lepton beams and polarised targets were used. The goal of these experiments was the measurement of the helicity distributions $\Delta q(x)$ (or $g_1^q(x)$), defined as difference between the number densities of quarks of flavor $q$ with helicity equal or opposite to that of the nucleon, for a nucleon polarised along its direction of motion (longitudinal polarisation). Following the discovery by the EMC at CERN in 1988 that the quark spins contribute only little to the proton spin, the interest in the nucleon spin structure was revived. In the past decade, a third generation of experiments at CERN (COMPASS), DESY (HERMES) and JLab began to study the nucleon spin structure using semi-inclusive DIS (SIDIS) processes, in particular the gluon contribution to the nucleon spin. These experiments also use polarised beams and polarised targets, while additionally hadrons produced by fragmentation of the struck quark are reconstructed and identified.

Since the 90's [1] it is well known that in order to fully specify the quark structure of the nucleon at twist-two level in quantum chromodynamics (QCD), the transversity distributions $\Delta_T q(x)$ (or $h_1^q(x)$) have to be introduced in addition to the momentum distributions $q(x)$ and the helicity distributions $\Delta q(x)$. For a given quark flavour $q$, $\Delta_T q(x)$ is the analog of the helicity distribution in the case of a transversely polarised nucleon. Helicity and transversity distributions coincide in the non-relativistic quark model but they are expected to be different when relativistic effects are taken into account. As there exists no gluon transversity distribution, the $Q^2$ evolution of $\Delta_T q(x)$ is quite different from that of $\Delta q(x)$. The first moment of the valence quark transversity distribution is related to the tensor charge $\delta q$,

$$\int_0^1 dx\,[\Delta_T q(x) - \Delta_T \bar{q}(x)] = \delta q\,, \tag{1}$$

which together with the vector and axial charge characterises the nucleon as a whole. The tensor charge is presently being calculated with steadily increasing accuracy by lattice QCD [2].

The transversity PDF is chiral-odd and thus not directly observable in inclusive deep inelastic lepton-nucleon scattering. In 1993 Collins suggested that it could be measured in SIDIS processes, where it appears coupled with another chiral–odd function [3], which by now is known as "Collins fragmentation function" $\Delta_T^0 D_q^h$ (or $H_{1q}^h$). It is the transverse-spin dependent part of the standard fragmentation function (FF) $D_q^h$ that describes the correlation of quark ($q$) transverse polarisation and hadron ($h$) transverse momentum. This mechanism leads to a left-right asymmetry in the distribution of hadrons produced in the fragmentation of transversely polarised quarks, which in SIDIS shows up as an azimuthal transverse spin asymmetry $A_{Coll}$ (the "Collins asymmetry") in the distribution of produced hadrons. At leading order this asymmetry can be written as

$$A_{Coll} = \frac{\sum_q e_q^2 \cdot \Delta_T q \otimes \Delta_T^0 D_q^h}{\sum_q e_q^2 \cdot q \otimes D_q^h}\,, \tag{2}$$

where $\otimes$ indicates the convolutions over transverse momenta. The Collins asymmetry is accessed through the amplitude of the $\sin \Phi_C$ modulation in the hadron azimuthal distribution. Here the Collins angle $\Phi_C = \phi_h + \phi_s - \pi$ is the sum of the azimuthal angles of the hadron transverse momentum $\vec{p}_T^{\,h}$ ($\phi_h$) and of the spin direction of the target nucleon ($\phi_s$) with respect to the lepton scattering plane, in a reference system in which the z axis is the virtual–photon direction.

A non-zero Collins asymmetry for the proton was first observed in 2004 by HERMES [4] using an electron beam of energy 27.6 GeV. These results provided first evidence that both Collins fragmentation and transversity functions are non-vanishing, although room was left to possible explanations of the observed signal in terms of higher–twist effects. Independent evidence of a non-zero and sizeable Collins function came soon after from measurements by the Belle Collaboration [5, 6] of the correlation between the azimuthal angles of hadrons in two jets resulting from $e^+e^-$ annihilations into hadrons, recently confirmed also by the Babar Collaboration [7]. For a comprehensive review of recent experiments and theoretical developments see e.g. refs. [8] and [9].

Using a 160 GeV $\mu^+$ beam COMPASS measured SIDIS on a transversely polarised $^6$LiD target in 2002, 2003 and 2004. In those data no size-able Collins asymmetry was observed within the accuracy of the measurements [10–12], a fact that was understood in terms of a cancellation between the u- and d-quark contributions [11]. The COMPASS data are still today the only SIDIS data ever taken on a transversely polarised deuteron target and provide constraints on the d-quark contribution [1]. Together with the HERMES and Belle data they allowed for the first global analyses and first extractions [14, 15] of the transversity distributions for u- and d–quarks and of favoured and unfavoured Collins FFs. It is important to note that global analyses of $e^+e^- \to hadrons$ and SIDIS data are a necessity, since neither can the Collins FF be extracted from $e^+e^-$ data alone, nor can the Collins and transversity functions be disentangled using only SIDIS data. Similar considerations hold for the transverse spin asymmetry in hadron pair production in DIS [16], where a signal on the proton was recently measured by HERMES [17] and COMPASS [18]. In this case the transversity distribution is coupled with the di-hadron FF which was recently accessed [19] using the Belle data.

In 2007 COMPASS measured for the first time SIDIS on a transversely polarised proton ($NH_3$) target using the $\mu^+$ beam of energy 160 GeV, thus extending the measured $x$ range to values about 10 times smaller as compared to HERMES. The results [20] for the Collins asymmetry in the valence quark region were in agreement with those of HERMES [21], in spite of the considerably larger $Q^2$ values.

As the Collins asymmetry includes convolutions of distribution and fragmentation functions depending on different variables, a large data sample is necessary in order to perform a multi-dimensional analysis and test the properties of the observable, namely factorisation and evolution. At COMPASS, the detailed investigation of the kinematic dependence of the Collins asymmetry was a strong motivation to improve on the precision of the 2007 measurement, and hence the entire 2010 run was dedicated to SIDIS measurements using the transversely polarised proton target ($NH_3$).

In this Letter, first results for the Collins asymmetry from the 2010 data are presented. The Sivers asymmetry, which is the amplitude of a $\sin(\phi_h - \phi_s)$ modulation in the hadron azimuthal distribution, was extracted from the same data and the results are given in a parallel paper [22].

The COMPASS spectrometer [23] is in operation in the SPS North Area of CERN since 2002. The principle of the measurements and the data analysis were already described in refs. [10–12, 20]. In 2010, the spectrometer configuration was very similar to that used in 2007 [20]. Additionally, a new triggering system for large-angle muons was used, which is based on two large area scintillator counter hodoscopes with 32 horizontal bars each and a suitable coincidence matrix to provide target pointing in the non-bending vertical plane. The target is polarised along the vertical direction and consists of three cylindrical cells with a diameter of 4 cm; the central cell is 60 cm long, and the two outer ones are 30 cm long and 5 cm apart. Neighbouring cells are polarised in opposite directions, so that data for both spin directions are recorded at the same time.

[1] The Collins asymmetry was very recently measured for the first time on a transversely polarised $^3$He target at JLab [13].

The $\mu^+$ beam had a nominal momentum of 160 GeV/c with a momentum spread $\Delta p/p = \pm 5\%$ and a longitudinal polarisation of -80%. The data were taken at a mean beam intensity of $3.5 \cdot 10^8$ $\mu$/spill, for a spill length of about 10 s every 40 s. About $37 \cdot 10^9$ events, corresponding to 1.9 PB of data, were collected in twelve separate periods. In order to minimise systematics, in each period after 4-5 days of data taking a polarisation reversal requiring 1.5 days was performed and then data taking continued for a corresponding number of days.

In the data analysis, in order to ensure the DIS regime, only events with photon virtuality $Q^2 > 1$ (GeV/c)$^2$, fractional energy of the virtual photon $0.1 < y < 0.9$, and mass of the hadronic final state system $W > 5$ GeV/c$^2$ are considered, leading to a total of $16 \cdot 10^7$ DIS events. The charged hadrons are required to have at least 0.1 GeV/c transverse momentum $p_T^h$ with respect to the virtual photon direction and a fraction of the available energy $z > 0.2$. After these cuts about $8 \cdot 10^7$ hadrons are left and used for the extraction of the asymmetries. This is referred to as "standard sample" in the following.

The asymmetries are measured separately for positive and negative hadrons as function of $x$, $z$ or $p_T^h$. The $\langle Q^2 \rangle$ values are about 3.7 (GeV/c)$^2$ in all $z$ and $p_T^h$ bins, while they strongly vary with $x$ due to the fixed target kinematics (see fig. 1). The mean values of $x$ in the $z$ and $p_T^h$ bins, of $z$ in the $x$ and $p_T^h$ bins, and of $p_T^h$ in the $x$ and $z$ bins are $\langle x \rangle = 0.05$, $\langle z \rangle = 0.38$, and $\langle p_T^h \rangle = 0.52$ GeV/c respectively.

In every bin of $x$, $z$ or $p_T^h$, for each period of data taking the asymmetries are extracted from the number of hadrons produced in each cell for the two directions of the target polarisation. Using an extended Unbinned Maximum Likelihood (UML) estimator [20], all 8 azimuthal modulations expected in the transverse spin dependent part of the SIDIS cross-section [24] are fitted simultaneously.

The measured amplitude of the modulation in $\sin\Phi_C$ is $\varepsilon_C = D_{NN} f P_T A_{Coll}$, where $D_{NN} = (1-y)/(1-y+y^2/2)$ is the transverse–spin–transfer coefficient from target quark to struck quark, $f$ the dilution factor of the $NH_3$ material, and $P_T$ the magnitude of the proton polarisation. In order to extract $A_{Coll}$, the measured amplitudes $\varepsilon_C$ in each period are divided by $f$, $P_T$ and $D_{NN}$. The dilution factor of the ammonia target is calculated for semi-inclusive reactions [25] and is evaluated in each $x$ bin; it increases with $x$ from 0.14 to 0.17, and it is assumed constant in $z$ and $p_T^h$. The proton target polarisation ($\sim 0.8$) was measured individually for each cell and each period. The final asymmetries are obtained by averaging the results of the 12 periods, after having verified their statistical compatibility.

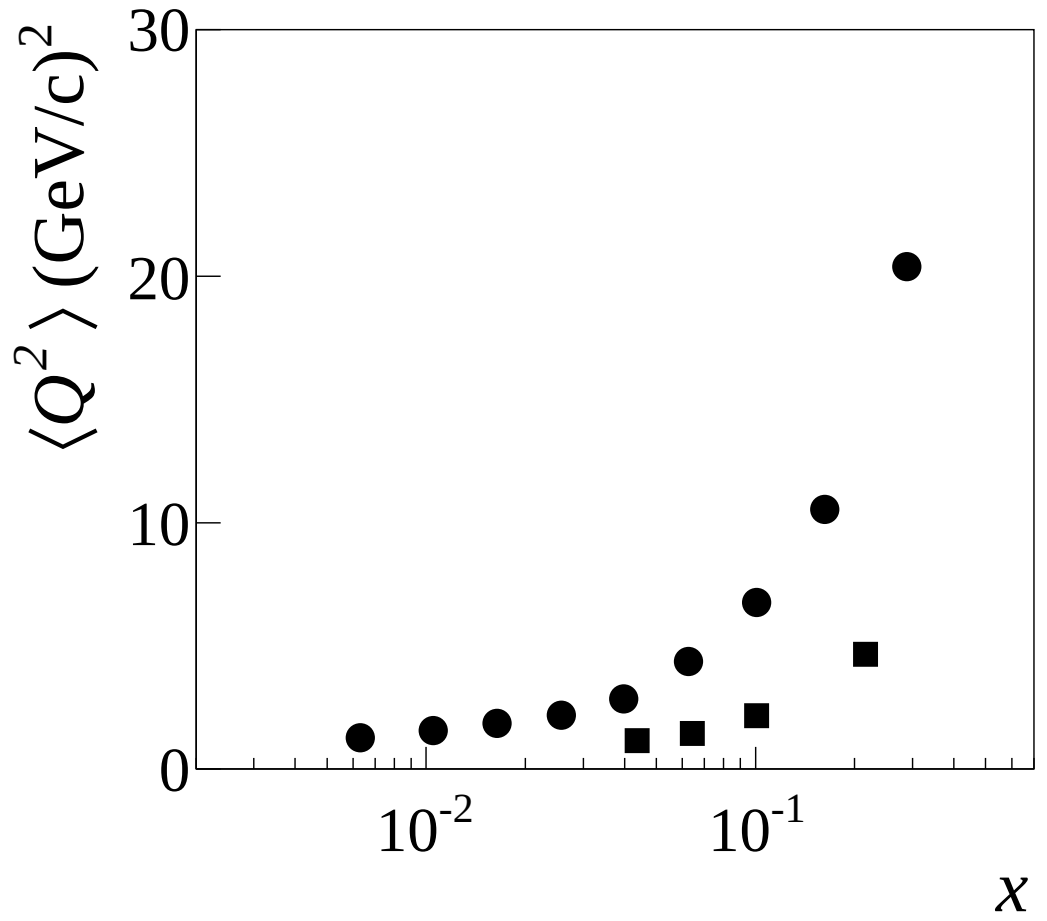

Fig. 1: Mean values of $Q^2$ in the $x$ bins for standard sample range $0.1 < y < 0.9$ (closed circles, •) and for events with $0.05 < y < 0.10$ (closed squares, ■).

Extensive studies were performed in order to assess the systematic uncertainties of the measured asymmetries, and it was found that the largest contribution is due to residual acceptance variations within the data taking periods. In order to quantify these effects, various types of false asymmetries are calculated from the final data sample assuming wrong sign polarisation for the target cells. Moreover, the physical asymmetries are extracted splitting the events according to the detection of the scattered muon in the spectrometer (top vs bottom, left vs right). The differences between these physical asymmetries and the false asymmetries are used to quantify the overall systematic point-to-point uncertainties, which are evaluated to be 0.5 times the statistical uncertainties. For further details on the evaluation of the systematic uncertainties of the 2010 data see ref. [26]. A systematic scale uncertainty of 3% arises from the measurement of the target polarisation.

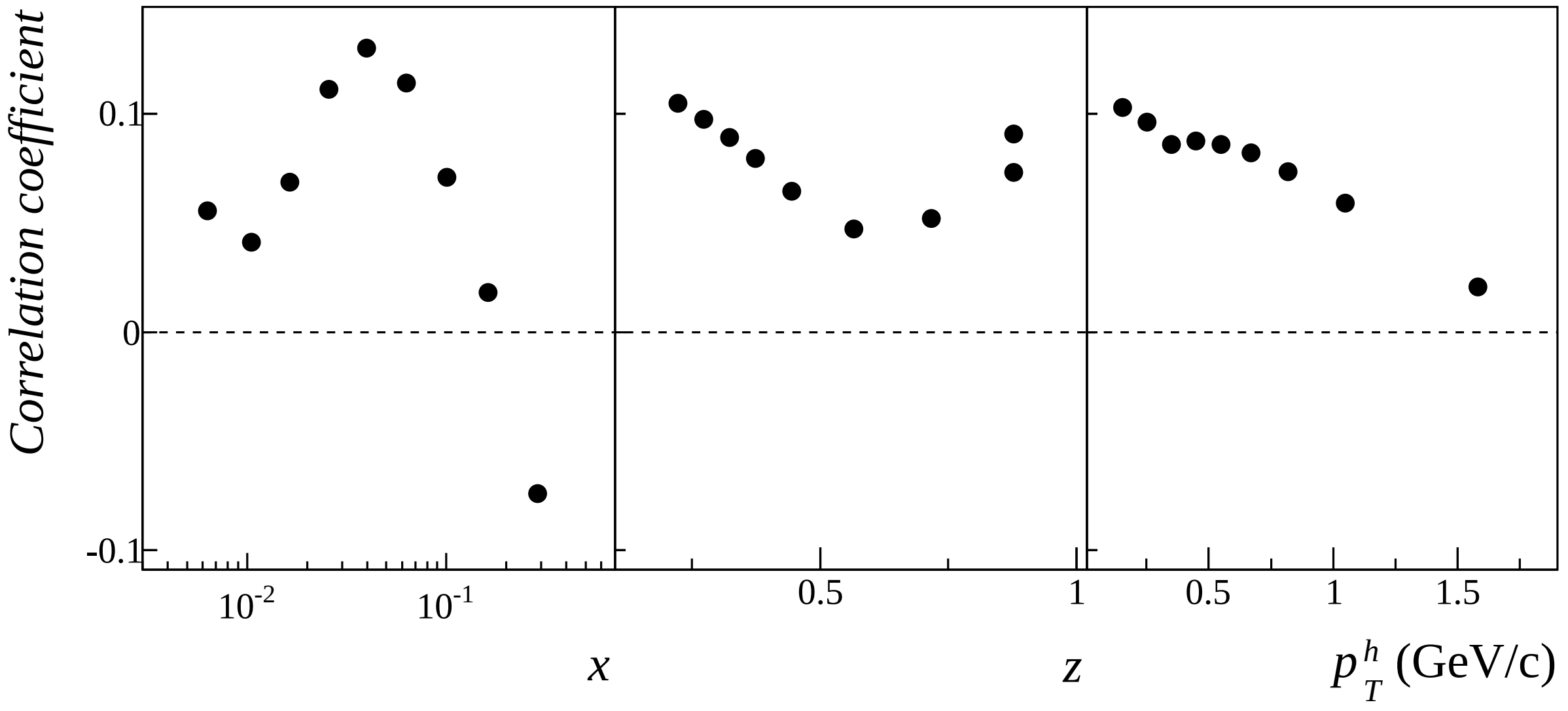

Fig. 2: Correlation coefficient between the Collins and Sivers asymmetries as a function of $x$, $z$ and $p_T^h$.

The correlations between the Collins asymmetries obtained after binning the data in $x$, in $z$ or in $p_T^h$ were also evaluated. The Collins asymmetries measured along different projections of the ($x$, $z$, $p_T^h$) phase space are statistically correlated, because their mean values have to be the same. In the COMPASS case these correlation coefficients are all smaller than about 0.2, but non-negligible, so that they should be taken into account in any global fit.

The correlations between the amplitudes of the 8 modulations allowed in the UML fit are found to be small and about the same for positive and negative hadrons. No correlations are expected a priori, since these azimuthal modulations are independent, but in the COMPASS experiment they arise mostly because the coverage in the horizontal plane of the muon triggering system is not complete. Figure 2 shows the correlation coefficients between the Collins and Sivers asymmetries as a function of $x$, $z$ and $p_T^h$, as given by the UML estimator. As can be seen, they are always smaller than 0.15. This correlation regards only the statistical fluctuations of the observables and is relevant only in case of simultaneous fits to the various asymmetries.

Figure 3 shows the Collins asymmetries measured as a function of $x$, $z$, and $p_T^h$ for positive and negative hadrons. The error bars are statistical only. As can be seen in the figure, the Collins asymmetry has a strong $x$ dependence. It is compatible with zero in the small $x$ region accessible at COMPASS and increases up to 0.05 in the valence quark region ($x > 0.1$), confirming that the transversity PDF is a valence object. The data exhibit a mirror symmetry with respect to the hadron charge, which is interpreted as due to the same size and opposite sign of the favoured and unfavoured Collins FFs [10]. The values are in agreement with our previous measurements [20], with an important gain in statistics: the statistical and the systematic uncertainties are reduced by about a factor of two. In the same figure, the data are compared with the predictions of ref. [27] which were obtained by fitting the HERMES proton data [28], the COMPASS deuteron data [12] and the Belle $e^+e^-$ fragmentation data [6]. The observed agreement supports the weak $Q^2$ dependence of the Collins asymmetry assumed in the calculation.

The high statistics of the 2010 data also allows for a thorough investigation of the kinematic dependences of the Collins asymmetry. As a starting point, the $z$ and $p_T^h$ dependences of the asymmetries are also evaluated for the region $x > 0.032$, where the signal is different from zero. In this region the mean value of $Q^2$ is equal to 5.75 (GeV/c)$^2$. As expected the results, given as open points in fig. 4, show a larger magnitude of the asymmetry due to the higher mean value of $x$, and are compatible with a linear dependence on $p_T^h$ up to about 1 GeV/c. These results agree in magnitude and sign with HERMES measurements [21] in this $x$ region that is common to both experiments, although the mean values of $Q^2$ in each $x$ bin are 3 to 4 times larger in COMPASS.

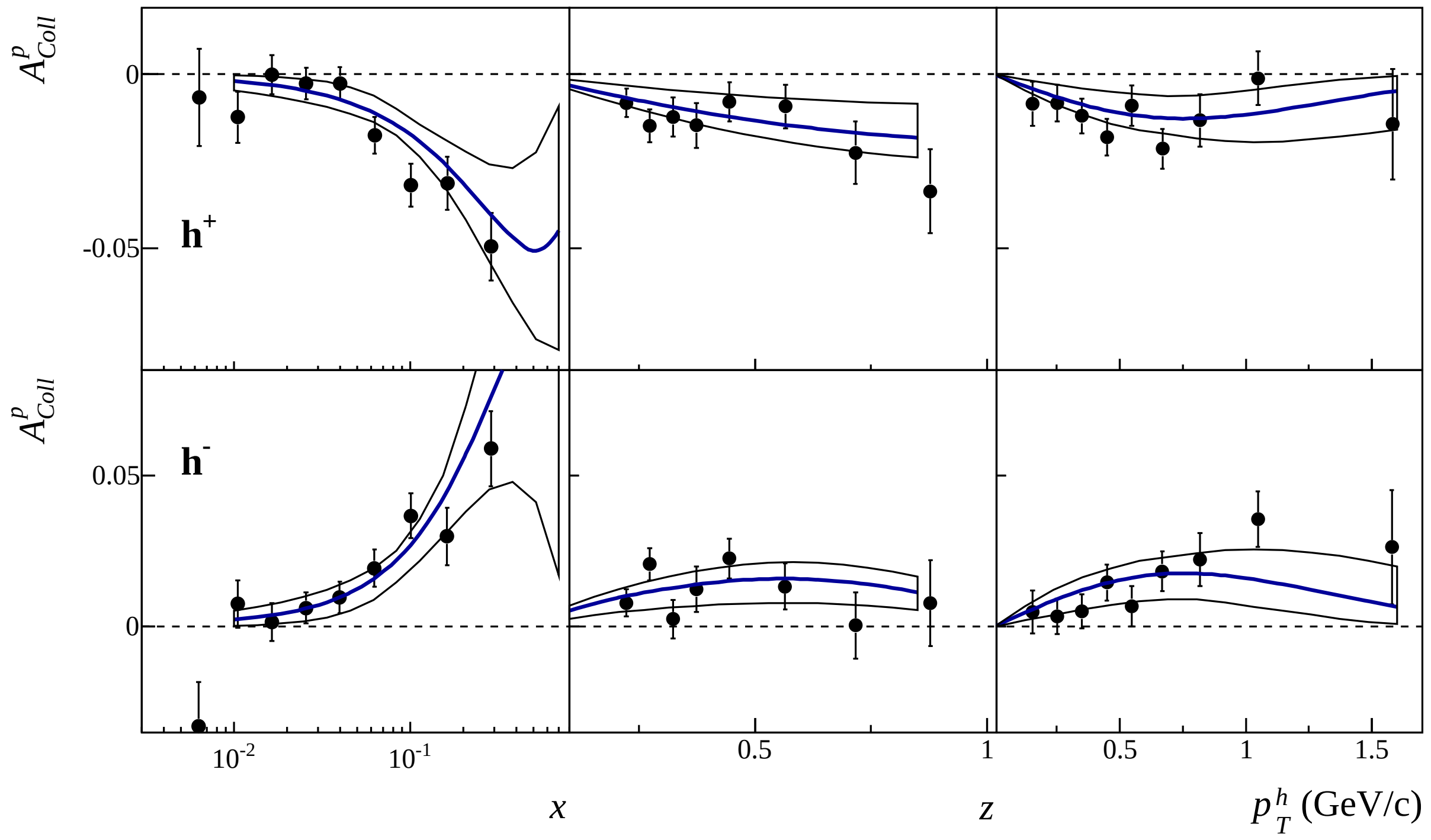

Fig. 3: Collins asymmetry as function of $x$, $z$ and $p_T^h$ for positive (top) and negative (bottom) hadrons. The curves are from ref. [27].

The Collins asymmetries are also studied at low $y$, namely in the region $0.05 < y < 0.1$. Given the strong correlation between $y$ and the mass of the hadronic final state system $W$, these events have $W$ values between 3 and 5 GeV/c$^2$, smaller than those of the standard sample. Also, as can be seen in fig. 1, for the $x$ range 0.032–0.70 $\langle Q^2 \rangle$ is smaller by about a factor of 3 with respect to the events with $0.1 < y < 0.9$. The measured Collins asymmetries for this low–$y$ sample are given in fig. 5, where they are compared with the results obtained by splitting the standard $y$ range into two bins, $0.1 < y < 0.2$ and $0.2 < y < 0.9$. The asymmetries measured as a function of $x$ are all compatible and only give some hint for a decrease at higher $y$ for negative hadrons. This is a further evidence that, if there is a $Q^2$ dependence of the Collins asymmetry, it has to be weak, which suggests both transversity and Collins functions being leading–twist quantities. Theoretical calculations of the $Q^2$ evolution of the transverse momentum dependent functions [29] are ongoing but not yet available for the Collins asymmetry.

The $z$ dependence of the Collins asymmetry is further studied extending the range towards the target fragmentation region, namely by measuring the asymmetries for hadrons with $0.1 < z < 0.2$ as a function of $x$ and $p_T^h$. The results are well compatible with those obtained from the standard $z > 0.2$ sample.

All the results given in this Letter are available on HEPDATA [30]. In particular, the asymmetries for the standard sample as functions of $x$, $z$ and $p_T^h$ have also been combined with the already published results

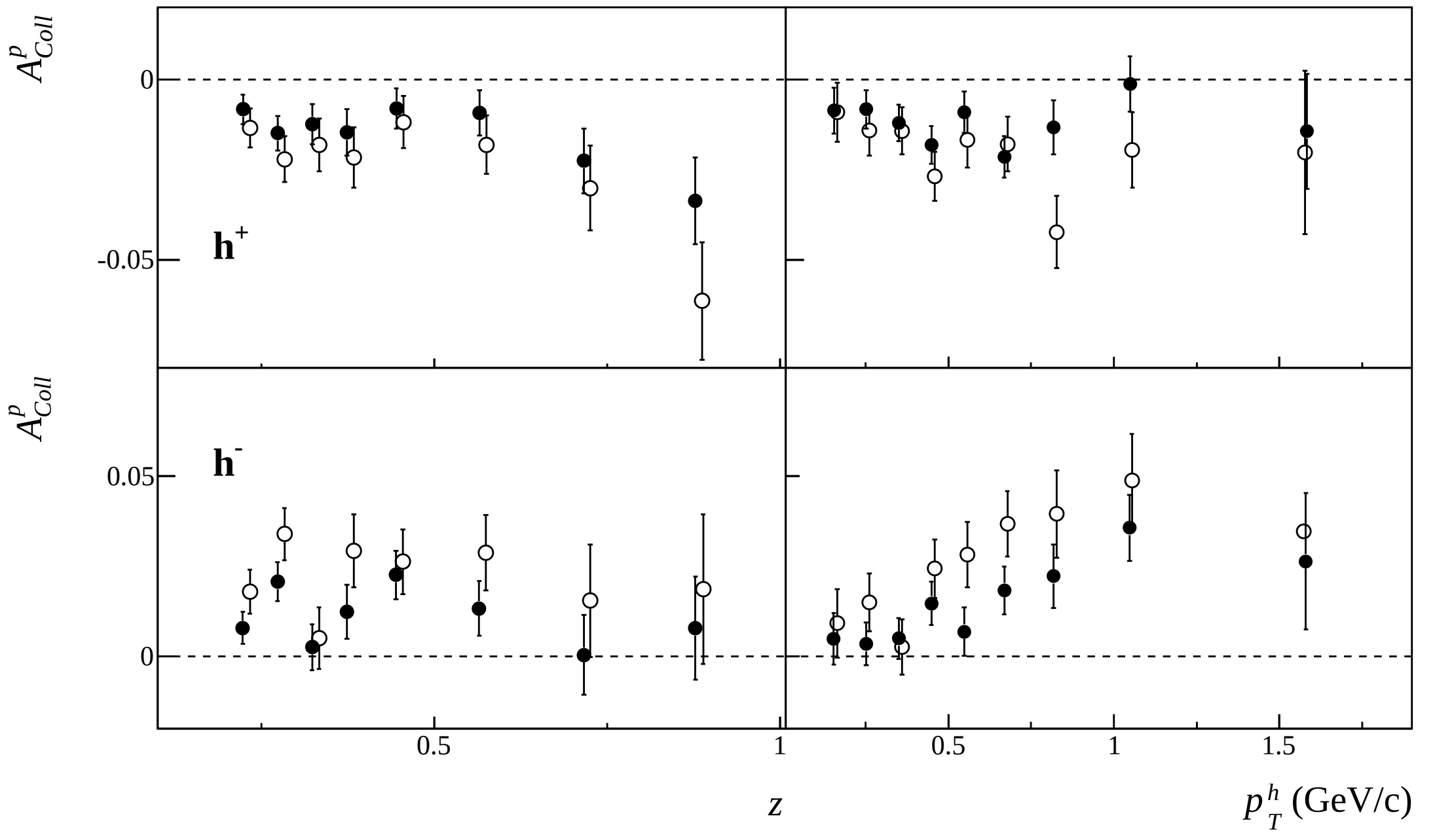

Fig. 4: Collins asymmetry as a function of $z$ and $p_T^h$ for positive (top) and negative (bottom) hadrons. The open points (○ , slightly shifted horizontally) are the values obtained in the range $x > 0.032$. The closed points (•) refer to the complete $x$ range and are the same as in fig. 3.

from the 2007 run [20] and are also available on HEPDATA.

In summary, this Letter presents the results from the 2010 COMPASS data for the Collins asymmetry on a proton target measured as a function of $x$, $z$ and $p_T^h$ using the entire collected statistics. They are compatible with the 2007 COMPASS results whereby improving the precision by a factor of two in the statistical and systematic uncertainties. The correlations between the Collins asymmetries measured along $x$, $z$ and $p_T^h$ respectively, and the correlations between the Collins and Sivers asymmetries have been investigated and found to be small. The Collins asymmetry has also been measured extending the kinematic range to smaller $z$ and $y$ values. The results confirm that the Collins asymmetry is a real leading–twist object. The availability of the precise COMPASS data on the proton, as well as the COMPASS data on deuteron, the HERMES proton results and the $e^+e^-$ fragmentation data, makes it mandatory to perform a new global analysis to disentangle the Collins and transversity functions and to study their properties.

We acknowledge the support of the CERN management and staff, as well as the skills and efforts of the technicians of the collaborating institutes.

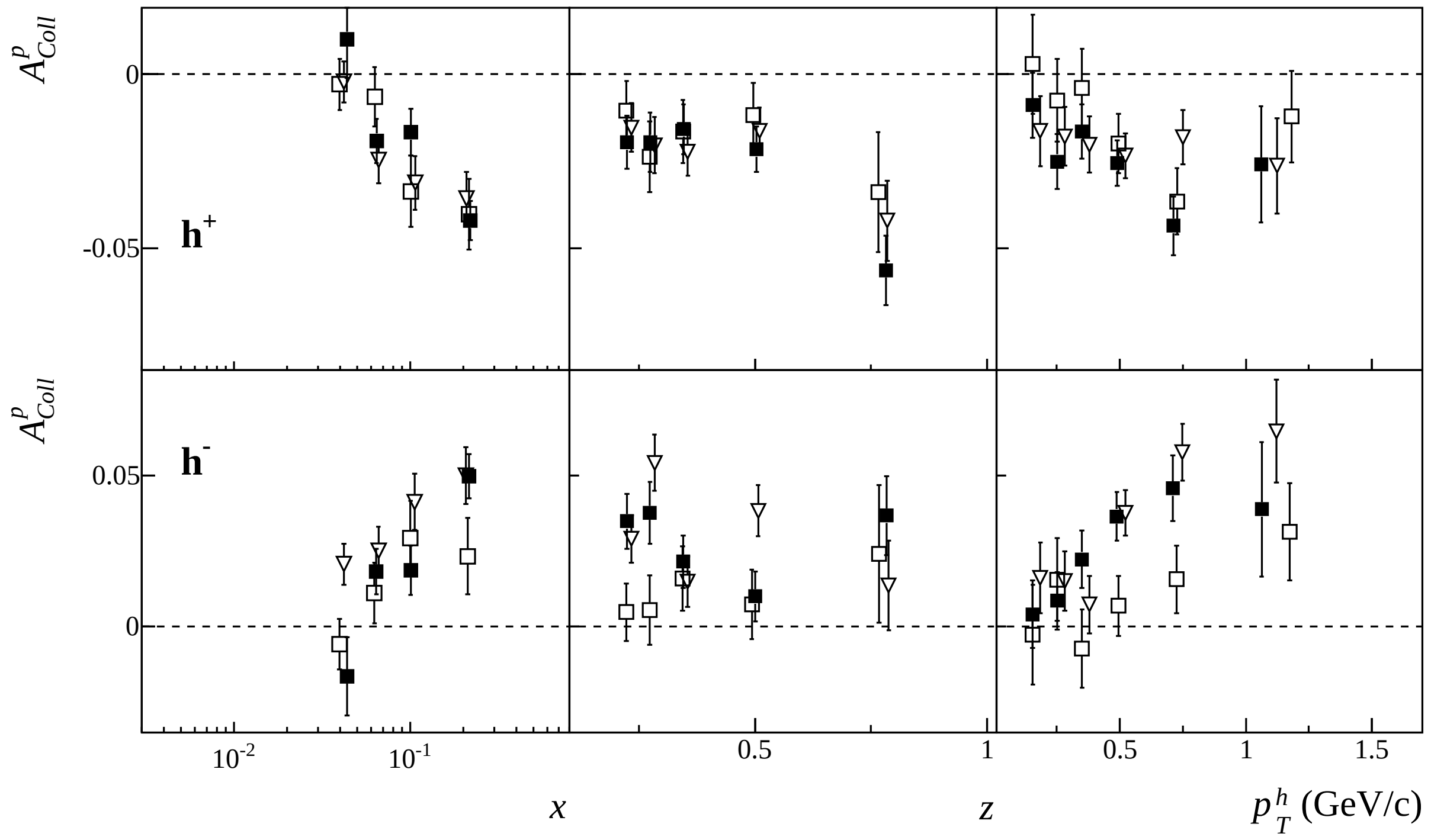

Fig. 5: Collins asymmetry as function of $x$, $z$ and $p_T^h$ for positive (top) and negative (bottom) hadrons in the range $x > 0.032$ and $0.05 < y < 0.1$ (closed squares, ■), $0.1 < y < 0.2$ (open triangles, ▽, slightly shifted horizontally) and $0.2 < y < 0.9$ (open squares, □).